# Stacked Charge Stripes in the Quasi-Two-Dimensional Trilayer Nickelate $La_4Ni_3O_8$

**Junjie Zhang[1,*], Yu-Sheng Chen[2], D. Phelan[1], Hong Zheng[1], M. R. Norman[1], and J. F. Mitchell[1,*]**

[1]Materials Science Division, Argonne National Laboratory, 9700 South Cass Avenue, Argonne, Illinois 60439, United States; [2]ChemMatCARS, University of Chicago, 9700 South Cass Avenue, Argonne, Illinois 60439, United States. [*]E-mail: junjie@anl.gov or mitchell@anl.gov

**Abstract:** The quasi-two-dimensional (quasi-2D) nickelate $La_4Ni_3O_8$ (La-438) is an anion deficient *n*=3 Ruddlesden-Popper (R-P) phase that consists of trilayer networks of square planar Ni ions, formally assigned as $Ni^{1+}$ and $Ni^{2+}$ in a 2:1 ratio. While previous studies on polycrystalline samples have identified a 105 K phase transition with a pronounced electronic and magnetic response but weak lattice character, no consensus on the origin of this transition has been reached. Here we show using synchrotron x-ray diffraction on high-$pO_2$ floating-zone grown single crystals that this transition is driven by a real space ordering of charge into a quasi-2D charge stripe ground state. The charge stripe superlattice propagation vector, $\boldsymbol{q}$=(⅔, 0, 1), corresponds with that found in the related ⅓-hole doped single layer R-P nickelate, $La_{5/3}Sr_{1/3}NiO_4$ (LSNO-⅓, $Ni^{2.33+}$) with orientation at 45° to the Ni-O bonds. The charge stripes in La-438 are weakly correlated along $\boldsymbol{c}$ to form a staggered *ABAB* stacking that minimizes the Coulomb repulsion among the stripes. Surprisingly, however, we find that the charge stripes within each trilayer of La-438 are stacked *in phase* from one layer to the next, at odds with any simple Coulomb repulsion argument.

**Introduction**

Competition between localized and itinerant electron behavior is an organizing construct in our understanding of correlated electron transition metal oxide (TMO) physics(1-4). Some of the most compelling phenomenology in these materials occurs in the mixed-valent state for the transition metal, which is set by composition, doping, and anion coordination of the metal. Many mixed-valent TMOs adopt insulating 'charge-ordered' states, in which an inhomogeneous yet long-range ordered configuration of the charge density condenses from a uniform metallic state(5). The real-space pattern of charge order varies by material(6-9), but a typically-observed motif is some variety of charge stripes. Such stripes have been observed in cobaltites(10-12), cuprates(13-15), nickelates(16-19), and manganites(20-22), albeit with highly materials dependent configurations that hinge on a balance among Coulomb, lattice, and magnetic exchange energies. For instance, charge stripes in layered nickelates typically stagger themselves from layer to layer to reduce the collective electrostatic energy arising from the charge disproportionation(9, 18).

Indeed, the case of nickelates plays a prominent role in charge-stripe physics(6-9, 16-19, 23-28), since mixed-valent $Ni^{2+}$ ($d^8$) and $Ni^{3+}$ ($d^7$) compounds such as $La_{2-x}Sr_xNiO_4$ (LSNO) are structurally and electronically related to high-$T_c$ superconductors and thus have been targeted as potential alternatives to the cuprates. Instead of superconductivity, however, the ground state of such quasi-two-dimensional (quasi-2D), octahedrally coordinated nickelates is marked by static charge and spin stripes, with a three-fold superlattice periodicity for compositions near $x=1/3$ ($La_{5/3}Sr_{1/3}NiO_4$, hereafter abbreviated as LSNO-⅓) found to be particularly stable(7). Despite this ubiquitous behavior for octahedrally coordinated nickelates, Anisimov *et al.* have suggested that $Ni^{1+}$ in a square planar coordination with O ions can form an $S=½$ antiferromagnetic insulator that may be doped with low spin ($S=0$) $Ni^{2+}$ holes to yield a superconductor(29).

To test such ideas, a series of anion-deficient Ruddlesden-Popper (R-P) phases $R_{n+1}Ni_nO_{2n+2}$ ($R$=La, $n$=2; $R$=La, Pr, Nd, $n$=3) with a rigorously square planar Ni-O environment were synthesized and studied in polycrystalline form(30-37). None of these materials superconducts. However, one of the members of the series, $La_4Ni_3O_8$ (La-438), undergoes a still incompletely understood phase transition on cooling through 105 K, accompanied by a dramatic increase in resistivity and a discontinuity in magnetization(35). Based on a range of theoretical treatments, the transition has been attributed to a spin-density wave (SDW)(35) or to a spin-state driven metal-insulator transition(37-39), possibly accompanied by charge disproportionation into two $Ni^{1+}$ sheets sandwiching a $Ni^{2+}$ sheet in the trilayer(Fig. 2*A*)(40). Difficulties remain with each of these potential explanations, and although NMR measurements(36) reveal strongly 2D antiferromagnetic spin fluctuations developing eventually into long-range order, the nature of the low temperature magnetic state remains open; neutron powder diffraction shows no magnetic peaks(35). Other fundamental issues connected to nickelate physics are also relevant to La-438, including homogeneity of the hole concentration in the symmetry-inequivalent layers (*i.e.*, layered charge segregation that must be present to some extent due to the different environments of the outer and inner layers), the appropriateness of a mixed-valent $Ni^{1+}/Ni^{2+}$ description, the role of ligand holes vis-à-vis more highly oxidized $Ni^{3+}$-containing oxides such as LSNO and $LaNiO_3$, $3z^2$-$r^2$/$x^2$-$y^2$ orbital polarization, and ultimately the potential for unconventional superconductivity(41). Unfortunately, a lack of single crystals to date has challenged definitive experimental tests and impeded progress on all of these issues.

Here, we present a fresh view of La-438 physics made possible by our newfound ability to grow single crystals of this compound. In particular, using single-crystal synchrotron x-ray diffraction, we find evidence of a superlattice below the 105 K transition and argue that it is driven by real-space ordering of charge. Our results reveal a close connection between La-438 ($Ni^{1.33+}$) and LSNO-⅓ ($Ni^{2.33+}$), including a shared propagation wave-vector, $q$=(⅔, 0, 1), ordering of the stripes at 45° to the Ni-O bonds, and a weak coupling along $c$ in a staggered configuration that minimizes Coulomb repulsion between adjacent trilayer blocks separated by ~6.5 Å. Remarkably, however, our data show that within the trilayer itself, the charge stripes are stacked *in phase* with one another, violating the Coulomb repulsion argument at this much shorter (~3.25 Å) length scale.

**Results and Discussion**

La-438 single crystals (1~2 mm² × 0.5 mm) were obtained by reducing (4% $H_2$/Ar gas, 350 °C, five days) specimens cleaved from a boule of $La_4Ni_3O_{10}$ that was grown at $pO_2$ = 20 bar in an optical-image floating zone furnace (HKZ-1, SciDre GmbH). The crystals are fragile, likely due to strains and microcracks that develop during the reduction process. Indeed, the structure undergoes a large, highly anisotropic expansion: $\Delta a \sim \Delta b \sim$ +3.0%, and $\Delta c \sim$ -6.6%)(34, 42). Nonetheless, La-438 specimens measured at 15-ID-B of the Advanced Photon Source (APS) definitively showed that they are well-defined single crystals, from which the structure of La-438 was determined (see *SI*).

At room temperature, La-438 crystallizes in the tetragonal *I*4/*mmm* space group with unit cell parameters $a$=3.9700(5) Å, $c$=26.092(3) Å and $Z$=2, in agreement with the structure reported by Poltavets *et al.* from Rietveld refinement on powder neutron data(34). For consistency with the body of literature on LSNO(18, 43), we henceforth adopt the *F*4/*mmm* ($\sqrt{2}a \times \sqrt{2}a \times c$) description of this high-*T* phase, with the principal axes rotated 45° from the Ni-O bonds (Fig. 2*A*). All Ni atoms are in square-planar coordination with Ni-O bond lengths 1.9850(2) Å for Ni(1)-O and 1.9852(3) Å for Ni(2)-O, which are slightly longer than those reported for the *n*=2 $La_3Ni_2O_6$(30) and *n*=∞ $LaNiO_2$(44) phases. Nominally, the average $Ni^{1.33+}$ oxidation state can be apportioned as $Ni^+$ and $Ni^{2+}$ in the outer and inner layers, respectively. However, first-principles theoretical considerations argue against this picture(39), as do our experimental data presented below.

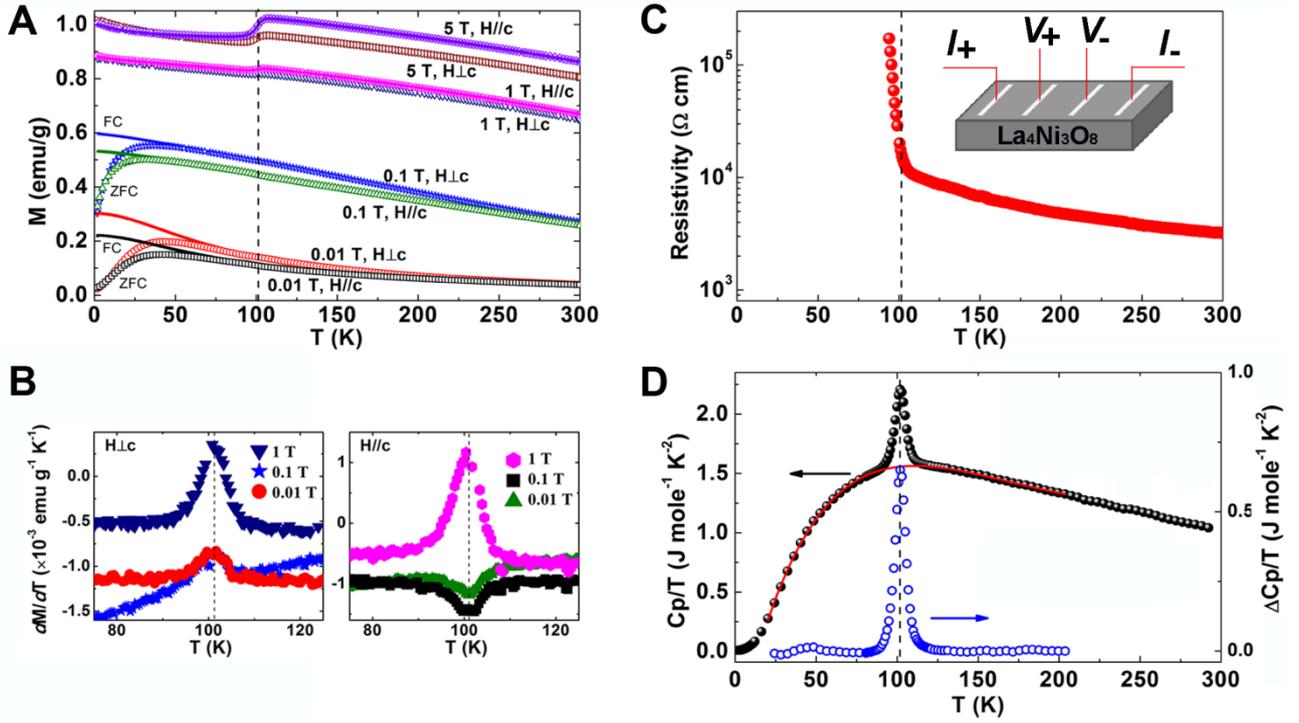

Fig. 1. Temperature dependence of selected physical properties of $La_4Ni_3O_8$. (*A*) In-plane and out-of-plane DC magnetic susceptibility for various magnetic fields (ZFC: zero field cooling, open symbols; FC: field cooling, solid symbols). (*B*) First derivative of in-plane and out-of-plane magnetic susceptibility for various magnetic fields. (*C*) Electronic resistivity in the *ab* plane. (*D*) Heat capacity in zero field. Solid back points: data; Red curve: polynomial background fit; Open blue circles: difference between the data and the fit.

Physical properties including magnetic susceptibility, resistivity, and heat capacity were measured on as-prepared La-438 single crystals(45), revealing anisotropy not available from powder specimens. Fig. 1*A* shows the temperature dependence of the in-plane ($\boldsymbol{H}\perp\boldsymbol{c}$) and out-of-plane ($\boldsymbol{H}\|\boldsymbol{c}$) DC magnetic susceptibility. Anomalies at 105 K are observed at both high and low applied magnetic fields. At high applied fields ($\mu_0 H \geq 1$ T), the magnetization shows an abrupt drop while cooling through *T*=105 K, in agreement with that reported by Poltavets *et al.*(35) and Cheng *et al.*(37). The decrease in *M* through the transition is field dependent and anisotropic. For example, the decrease in *M* is 6.5%($\boldsymbol{H}\|\boldsymbol{c}$) and 2.5%($\boldsymbol{H}\perp\boldsymbol{c}$) at 5 T. This anisotropy increases at low fields, reaching ~7 at 1 T, accompanied by a less pronounced anomaly. At low applied fields ($\mu_0 H \leq$ 0.1 T), *M* shows a subtle, anisotropic change in slope that was apparently obscured in the polycrystalline samples measured by Poltavets *et al.*(35). To emphasize this behavior, *dM/dT* for low fields is presented in Fig. 1*B*. The derivative for $\boldsymbol{H}\perp\boldsymbol{c}$ shows a maximum at each field, while that for $\boldsymbol{H}\|\boldsymbol{c}$ shows a minimum at low fields and maximum at high fields. Similar behavior found in other layered oxides such as $Ca_3Co_4O_9$(46) and $Na_{0.71}CoO_2$(47) has been attributed to SDW formation. One possibility, then, is that the ground state of La-438 is an SDW, which would be consistent with the proposal of Poltavets *et al.* based on band structure calculations(35).

Another possibility is that the ground state magnetism develops out of a charge ordered precursor phase such as that found in related R-P nickelates(16-19). For example, LSNO-⅓, which charge orders at $T_{CO}$ ~239 K before spin order is established at $T_{SO}$ ~ 194 K, also exhibits a similar step in the magnetic susceptibility(48). Fig. 1*C* shows the temperature-dependent resistivity of the La-438 single crystals in the *ab* plane. It reveals a semiconducting behavior above the 105 K transition(49) and highly insulating behavior below this transition. Our data are consistent with those of Poltavets *et al.*(35) and Cheng *et al.*(37), and confirm an abrupt localization of charge below 105 K, qualitatively similar to that observed in LSNO-⅓(16). We note that Cheng *et al.*(37) suggest that poor grain connectivity in their cold-pressed polycrystalline La-438 samples leads to an anomalously high resistance for $T > 105$ K and argue from thermopower measurements that the true state in this high-$T$ regime is metallic. Our data show a similar high resistance, which could result from poor connectivity across a strain-induced network of microcracks created during the reduction process. Nonetheless, the temperature dependence is not that of a metal.

Fig. 1*D* shows the heat capacity of a La-438 single crystal plotted as $C_p/T$. To estimate the entropy change of the 105 K transition, we have phenomenologically fit the behavior above and below with a fifth-order polynomial and subtracted this background(50). Integrating the area under the resultant peak yields $\Delta S = 5.93$ J mole$^{-1}$ K$^{-1}$, which agrees well with that found by Poltavets *et al.* (5.96 J mole$^{-1}$ K$^{-1}$)(35). We note that the entropy change per Ni (1.98 J mole$^{-1}$ K$^{-1}$) in La-438 is close to that found (2.0±0.3 J mole$^{-1}$ K$^{-1}$) by Klingeler *et al.* in LSNO-⅓, which has been attributed to condensation of short-range, fluctuating charge stripes(48).

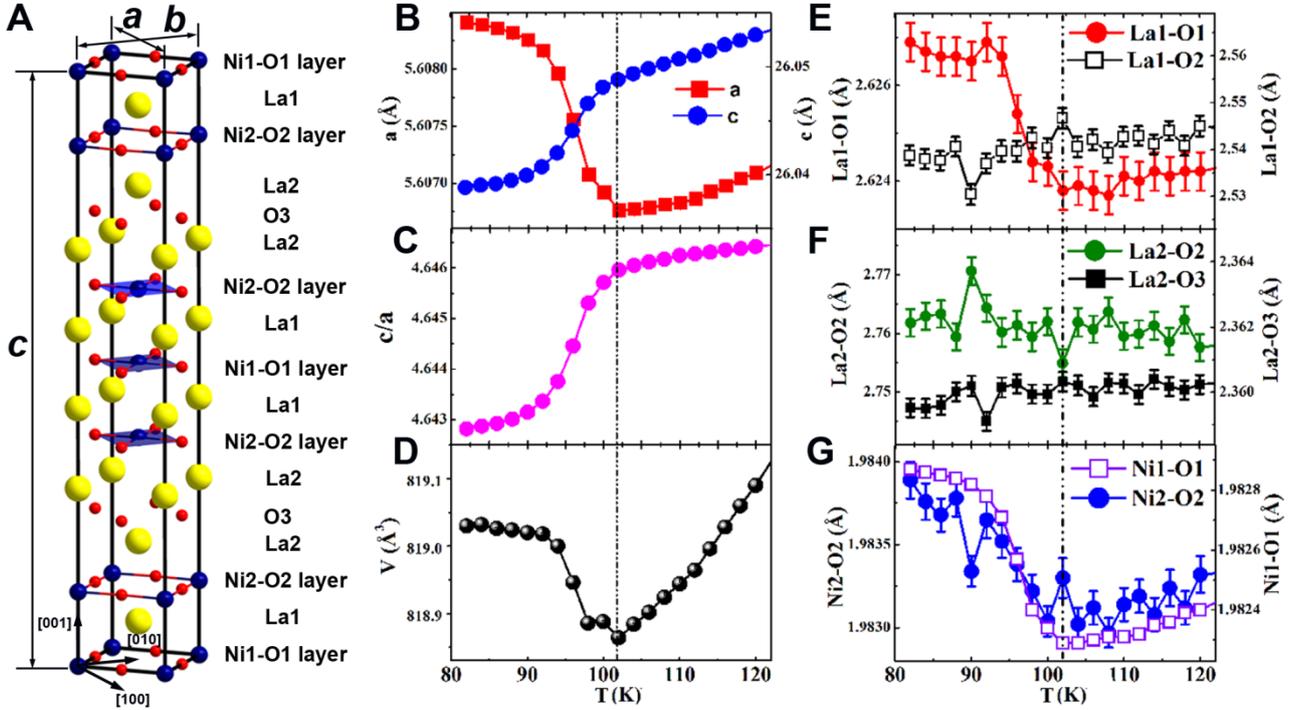

Fig. 2. Crystal structure of $La_4Ni_3O_8$. (*A*) The *I4/mmm* unit cell is highlighted with solid lines, the *F4/mmm* setting used in this work is identified by the lattice vectors ***a***, ***b***, and ***c***. Temperature dependence of the unit cell parameters (*B-D*) and bond distances (*E-G*) extracted from Rietveld refinement on high-resolution synchrotron X-ray powder diffraction data. Error bars represent the estimated standard deviation from the refinement, and they are smaller than symbols for the unit cell parameters (*a*, *c* and *V*).

Fig. 2*B-D* show the temperature dependence of the unit cell parameters in the range 82-120 K extracted from Rietveld refinement of high resolution x-ray powder diffraction data (see *SI*). No symmetry change was observed below the transition in this temperature range. However, a clear

increase of the *a* axis length (~0.029%) and unit cell volume (~0.02%), and a drop of the *c* axis length (~0.038%) are observed when cooling through the 105 K transition, with a concomitant drop in *c/a* (~0.068%). Our data are consistent with those of Cheng *et al.*(37), and reflect a weak lattice contribution to the transition. Fig. 2*E-G* show the La-O and Ni-O bond lengths as a function of temperature. Pronounced but smooth changes in the bond lengths of La1-O1(~0.003 Å), Ni1-O1 (<0.001 Å), and Ni2-O2 (<0.001 Å) are observed, but no changes outside the noise are observed in La1-O2, La2-O2 and La2-O3. This demonstrates that the structural changes that occur through the transition are isolated to the Ni-O trilayer blocks, with the rocksalt LaO layers acting as weak structural links between these blocks. We note that the sign of the change in unit cell parameters and Ni-O bond distance on cooling is consistent with the spin-state transition model proposed by K. Lokshin *et al.*(38) and developed theoretically by Pardo and Pickett(39), in which $x^2$-$y^2$ orbitals become more electron-rich. If such a picture is correct, however, the charge redistribution is associated with a remarkably small magnitude of the lattice response.

With an average oxidation state of 1.33+, square-planar coordinated La-438 is separated by an integral charge from the average oxidation state of 2.33+ in octahedrally coordinated LSNO-⅓, which develops a charge- and spin-stripe ordered ground state. It is thus intriguing to consider that analogous ordering of charges and spins could be occurring in the present square-planar coordinated trilayer system, electronically controlled by a ⅓-hole doping beyond a uniform $Ni^{1+}$ background. We now provide synchrotron x-ray diffraction evidence that supports such a picture, and argue that real space ordering of charge is likely to be the primary driving force of the 105 K transition in La-438.

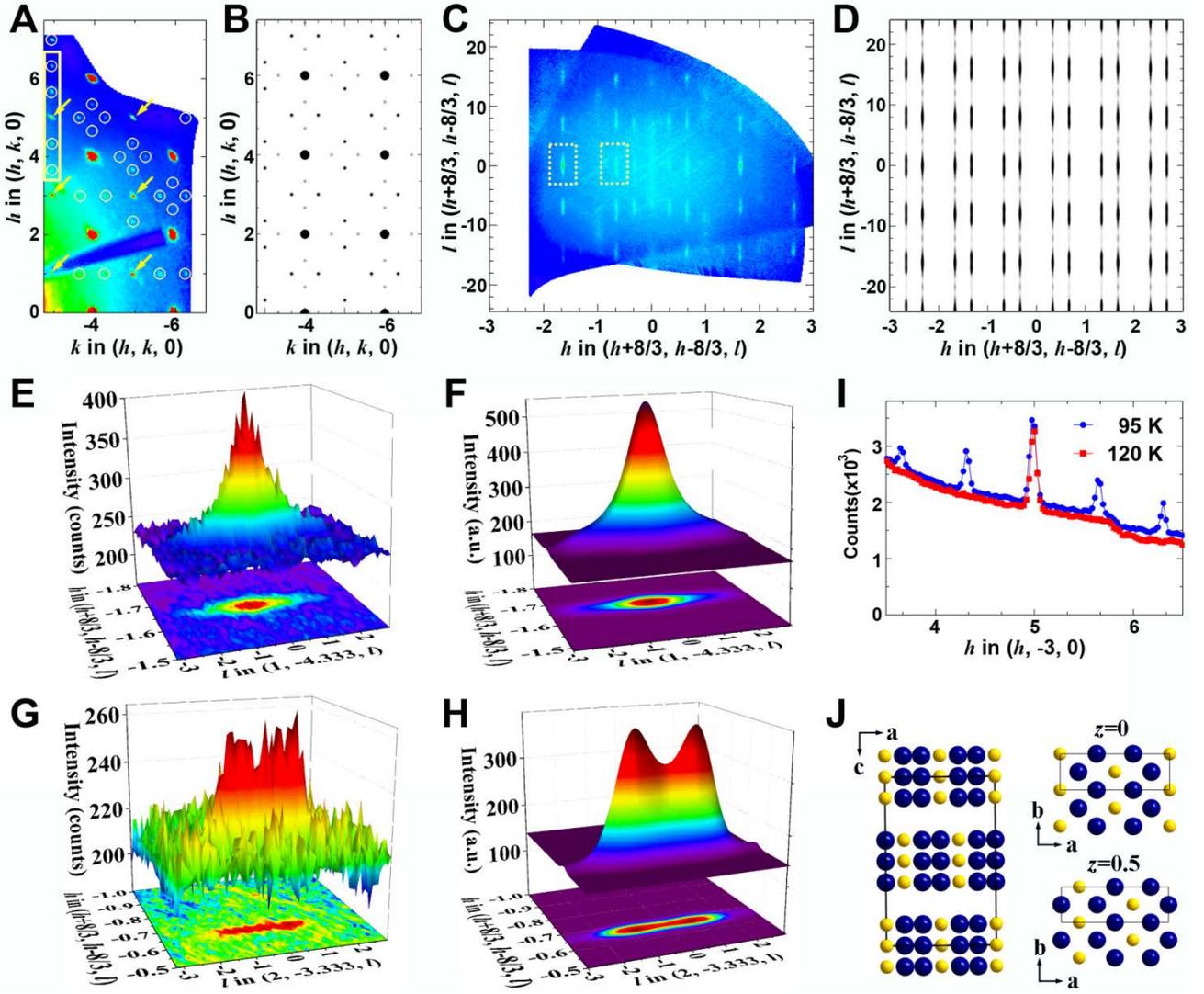

Fig. 3. Charge stripe ordering in $La_4Ni_3O_8$. (*A*, *C*) Reconstructed *hk*0 and (*h*+8/3, *h*-8/3, *l*) planes (*F*4/*mmm* notation) measured at 95 K at 15-ID-B. Note that intensity at the forbidden peaks (*e.g.* $1\bar{5}0$, $3\bar{5}0$, $5\bar{3}0$, ...), as indicated by yellow arrows in (*A*), result from diffuse rods of scattering along the *l* direction (see *SI* for details). (*E*, *G*) Measured superlattice peaks at (1, -4.333, 0) and (2, -3.333, 0), as indicated by dash yellow rectangle in (*C*). (*B*, *D*, *F*, *H*) Simulated diffraction patterns based on the charge stripe model shown in (*J*). The size of the spots in (*B*) indicates the intensity. (*I*) Intensity along the line (*h*, -3, 0) at 95 K (blue circles) and 120 K (red squares). (*J*) Charge stripe model in real space. Yellow(blue) spheres represent areas of higher(lower) valence than the average 1.33+. Solid rectangle denotes the charge stripe supercell.

Fig. 3*A* shows an *hk*0 plane measured at 95 K. In addition to the fundamentals at $2\bar{4}0$, $4\bar{4}0$, $2\bar{6}0$, *etc.*, sharp superlattice (SL) reflections are observed between these fundamentals, spaced evenly at an interval of ***a****/3. Refinement of the positions against ~400 SL reflections ($I \geq 3\sigma$) gives ***q***=0.333(1). SL spots are observed both along ***a**** and ***b****, reflecting the mixture of two 90° oriented orthorhombic domains expected upon lowering of symmetry from tetragonal to orthorhombic. The resulting pattern in the *hk*0 layer thus appears as squares of SL peaks along the zone edge, corresponding to the [100] direction in real space, or 45° to the Ni-O bonds. A line cut along (*h*, -3, 0), Fig. 3*I*, shows that these SL reflections are absent above the 105 K transition. In contrast to the sharp reflections seen in the *hk*0 projection, the SL peaks become broadened along $c^*$, as shown in the ($h$+8/3, $h$-8/3, $l$) plane (Fig. 3*C*). Unfortunately, the data are of insufficient quality to extract correlation lengths reliably; however, the qualitative appearance is consistent with the expected $\xi_{ab} \gg \xi_c$, as likewise found in LSNO-⅓(19). We might expect that short-range, fluctuating stripes will form well above 105 K before condensing, again like LSNO-⅓(23). The close agreement of the entropy change through the transition to long-range order between La-438 and LSNO-⅓ corroborates this supposition.

The distribution of positions and intensities of the SL reflections provides a means to assess the real space arrangement of the stripes, *i.e.* the intra- and interlayer charge stacking. Consider first the stacking of charges within the trilayer, a unique feature of La-438 vis-à-vis LSNO-⅓. Although Coulomb repulsion arguments favor a staggering of the charge from layer to layer, our diffraction data conclusively show that the charges are stacked *in phase* within the trilayer. Fig. 3*B* and *D* show simulations of the scattering from a stacked charge trilayer model assuming a propagation vector $\boldsymbol{q}$=(⅔, 0, 1). This produces a three-fold diagonal stripe pattern in the *ab*-plane oriented at 45° to the Ni-O bonds, in agreement with the observations. Intensity variation along $c^*$ takes maximum values at positions spaced by $\Delta l$=8, a consequence of the $c$/8 spacing of the layers within a trilayer. Importantly, the maxima along $c^*$ appear at $l$=8$n$, in agreement with the data presented in Fig. 3*C*. In contrast, a model in which the charges are staggered within the trilayer yields intensity maxima at $l$=4+8$n$ (see Fig. S3), inconsistent with the experimental data. Other stacking models were tested (see *SI*) but likewise failed to match the observed scattering. Stacking of charge is counterintuitive within a framework of a simple electrostatic model. Indeed, charge stacking in manganites adoping the zig-zag CE charge, orbital, and magnetic ordered state was explained by invoking an intersite Coulomb repulsion of lesser importance than the kinetic energy (double exchange) and magnetic interactions along the $Mn^{3+}/Mn^{4+}$ stripes(21). In the case of the La-438 nickelate, a similar mechanism might apply. A free energy gain is possible from the "$d^8$" sites if they stack on top of one another, because their partially occupied $3z^2$-$r^2$ orbitals would generate a net bonding interaction along $\boldsymbol{c}$. A related bonding picture was suggested by Pardo and Pickett(51), although their particular model should not exhibit the charge modulation found here.

We turn now to the inter-trilayer stacking of charges, which in single layer LSNO-⅓ follows a body-centering pattern schematically shown in Fig. 3 of Ref. (9) and Fig.10(c) of Ref. (18). This arrangement minimizes the Coulomb repulsion among the stripes. Our data indicate that La-438 adopts a competing *ABAB* stacking (*A* and *B* represent a trilayer) shown schematically in Fig. 3*J*. Fig 3*A* shows that the SL spots in the *hk*0 plane have a nonuniform intensity distribution, with reflections ($h$±⅓, $k$±1, 0) (and equivalents for the 90° rotated domain), where ($h$, $k$, 0) is an allowed reflection of *F*4/*mmm*, appearing more intense than those with index ($h$, $k$±⅔, 0). This distribution of intensity can be reproduced (Fig. 3*B*) by a model with *ABAB* stacking of the stripes along $\boldsymbol{c}$, akin to that of Fig. 10(b) of Ref. (18). Note that this *ABAB* stacking sequence permits even-$l$ reflections and thus differs from the LSNO-⅓ stacking pattern in which bond- and site-centered stripes alternate along $\boldsymbol{c}$, and for which odd-$l$ reflections only are observed.

Additional support for the *ABAB* model can be found by interrogating the $l$-dependence of the peak intensities. Fig. 3*E* and *G* are intensity plots along $c^*$ in the vicinity of the (1, -4.333, 0) and (2, -3.333, 0) SL reflections, respectively. While both are broad, there is a clear qualitative difference in the appearance of these peaks: the former consists of a single maximum, while the latter appears as a poorly resolved doublet. The simulated patterns for these reflections are shown in Fig. 3*F* and *H*, and are in good agreement with the corresponding data. Inspection of the simulated data in the absence of broadening shows that each peak is actually a triplet with intensity centered at *l*= -1, 0, 1. In the case of (1, -4.333, 0) and its equivalent SL reflections, the maximum intensity is located at *l*=0, while for (2, -3.333, 0), the maximum intensity is located at *l*= -1 and 1. Hücker *et al.*(18) have pointed out that the difference between a double period *ABAB* and a triple period *ABCABC* stacking is set by the second-neighbor interaction, which is expected to be weak. Due to the breadth of the SL reflections along *c\**, the possibility that La-438 has some *ABC* stacking faults cannot be ruled out entirely; however, the presence of *l* even SL reflections demonstrates that *ABAB* stacking of the trilayer blocks dominates. Thus, while La-438 adopts a different stacking than that found in LSNO-⅓, the stacking of charge from trilayer to trilayer in the observed pattern likewise minimizes the Coulomb repulsion.

**Conclusion**

Our observations offer a new perspective on La-438(35, 40, 51-54). Primarily, we find that the low-temperature ground state of La-438 is a charge-stripe ordered insulator, akin to those found in LSNO-⅓, and that this real-space order into charge stripes is the primary driver of the 105 K phase transition. We find that this inhomogeneous charge distribution takes the form of stacked charge stripes within the trilayer, which then stagger between trilayers along *c* in an *ABAB* sequence. Thus, when framed solely within the context of a simple inverse-square law, La-438 adopts a paradoxical ground state in which the Coulomb interaction is minimized at long distances but not at short-range. While a full, quantitative understanding of this stripe organization will require theoretical input, one possibility is that bonding among partially occupied $3z^2$-$r^2$ orbitals within the trilayer, which favors locking the charge into this stacked pattern, outweighs the short range electrostatic energy.

Further confirmation of the connection between La-438 and LSNO nickelate physics will require understanding the magnetic ground state of La-438, which should likewise exhibit the same in plane wave-vector; neutron and magnetic x-ray scattering measurements are planned to test this. Finally, our single crystals provide a platform to address a broader set of questions about La-438, including orbital polarization and related questions regarding spin-state and ligand-hole contribution. It may also be possible to dope electrons into these crystals to explore the impact on the charge ordered ground state.

**Materials and Methods**

**Physical Properties.** Magnetic susceptibility measurements were performed on single crystals using a Quantum Design MPMS3 SQUID magnetometer. Zero-field cooled and field cooled data in the *ab* plane and out-of-plane were collected between 1.8 and 300 K under an external field of 0.01, 0.1, 1.0, and 5.0 T. The electrical resistivity was measured in the *ab* plane of the $La_4Ni_3O_8$ single crystals under zero applied magnetic field using the standard four-probe AC technique on a Quantum Design Physical Properties Measurement System (PPMS). The specific heat measurements were carried out in the PPMS using the relaxation method under zero applied magnetic field.

**X-ray Diffraction Experiments**. Single crystal X-ray diffraction data were collected with an APEX2 area detector using synchrotron radiation ($\lambda$=0.41328 Å) at Beamline 15-ID-B at the Advanced Photon Source, Argonne National Laboratory. A single crystal with dimensions of approximately 5 μm on an edge was used to determine the structure at room temperature. To observe the superlattice peaks below the 105 K transition clearly, larger single crystals (approximately 20 μm on edge) and longer exposure times (*e.g.*, 2.0 s/0.2°) were used. In this case, many Bragg peaks were found to overflow due to the limited dynamic range of the CCD detector (max. 65536) and the extreme intensity ratio of the Bragg/superlattice peaks ($\sim 10^5$). Several single crystals were used in this experiment, and the superlattice peaks are reproducible. Data were collected at room temperature first, and then the samples were measured in the range of 90-200 K by flowing nitrogen gas, and in the range of 15-70 K by flowing helium gas. $\Phi$-scans were used, and 1800 frames were collected for *q*-vector refinements.

**Variable-temperature High Resolution Powder X-ray Diffraction**. High-resolution x-ray powder diffraction data were collected on pulverized La-438 crystals at beamline 11-BM (APS) in the range of $0.5° \leq 2\theta \leq 36°$ with a step size of 0.001° and step time 0.2 s (see *SI*).

**Stripe Ordering Simulations**. The structure factors for superlattice reflections were calculated via construction of superlattice unit cells with a modulation of charge imposed upon the nickel sites such that a charge disproportionation results. Although only an approximation to the true charge state of the stripes, this approach allows for a qualitative comparison with the observed pattern of superlattice reflections and thus provides insight into which models work and which can be eliminated. We have convolved the Bragg peaks with a Gaussian along the $c^*$ direction ($\sigma=0.5|c^*|$) to approximate the breadth of the observed peaks along the $c^*$ direction. The following simplifications have been employed: a real scattering length has been considered for each atom; no effort has been made to correct for the energy-dependent or momentum-dependent scattering length; Debye-Waller factors were not incorporated (see *SI*).

**ACKNOWLEDGEMENTS**. This work was supported by the U.S. Department of Energy, Office of Science, Basic Energy Sciences, Materials Science and Engineering Division. ChemMatCARS Sector 15 is supported by the National Science Foundation under grant number NSF/CHE-1346572. Use of the Advanced Photon Source at Argonne National Laboratory was supported by the U.S. Department of Energy, Office of Science, Office of Basic Energy Sciences, under Contract No. DE-AC02-06CH11357. The authors thank Mr. Wenyang Gao for his help with the synchrotron x-ray single crystal measurements at 15-ID-B, Dr. Saul Lapidus for his help with the high resolution x-ray powder diffraction at 11-BM, and Drs. V. Pardo, W. E. Pickett, J. W. Freeland, S. Rosenkranz, A. S. Botana, Y. Ren and C. D. Malliakas for helpful discussions.

**Author contributions**: J.F.M. and J.Z. designed research; J.Z., D.P., H.Z., and Y.S.C. performed research; J.Z., D.P., J.F.M., M.R.N, and Y.S.C. analyzed data; J.Z. and J.F.M. wrote the paper with contributions from all authors.

**Competing Interests**: The authors declare no conflict of interest.

**References**

1. Goodenough JB. Electronic and ionic transport properties and other physical aspects of perovskites. *Rep. Prog. Phys.* 67(11):1915-1993 (2004).
2. Imada M, Fujimori A, Tokura Y. Metal-insulator transitions. *Rev. Mod. Phys.* 70(4):1039-1263 (1998).
3. Austin IG, Mott NF. Metallic and nonmetallic behavior in transition metal oxides. *Science* 168(3927):71-77 (1970).
4. Goodenough JB. Perspective on engineering transition-metal oxides. *Chem. Mater.* 26(1):820-829 (2014).
5. Coey M. Charge-ordering in oxides. *Nature* 430(6996):155-157 (2004).
6. Ulbrich H, Braden M. Neutron scattering studies on stripe phases in non-cuprate materials. *Physica C* 481:31-45 (2012).
7. Tranquada JM. Spins, stripes, and superconductivity in hole-doped cuprates. *AIP Conf. Proc.* 1550(1):114-187 (2013).
8. Zaliznyak IA, Tranquada JM, Gu G, Erwin RW, Moritomo Y. Universal features of charge and spin order in a half-doped layered perovskite. *J. Appl. Phys.* 95(11):7369-7371 (2004).
9. Tranquada JM. Stripe correlations of spins and holes in cuprates and nickelates. *Ferroelectrics* 177(1):43-57 (1996).
10. Boothroyd AT, Babkevich P, Prabhakaran D, Freeman PG. An hour-glass magnetic spectrum in an insulating, hole-doped antiferromagnet. *Nature* 471(7338):341-344 (2011).
11. Cwik M, *et al.* Magnetic correlations in $La_{2-x}Sr_xCoO_4$ studied by neutron scattering: possible evidence for stripe phases. *Phys. Rev. Lett.* 102(5):057201 (2009).
12. Sakiyama N, Zaliznyak IA, Lee SH, Mitsui Y, Yoshizawa H. Doping-dependent charge and spin superstructures in layered cobalt perovskites. *Phys. Rev. B* 78(18):180406 (2008).
13. Comin R, *et al.* Broken translational and rotational symmetry via charge stripe order in underdoped $YBa_2Cu_3O_{6+y}$. *Science* 347(6228):1335-1339 (2015).
14. Abbamonte P, *et al.* Spatially modulated 'Mottness' in $La_{2-x}Ba_xCuO_4$. *Nature Phys.* 1(3):155-158 (2005).
15. Vojta M, Vojta T, Kaul RK. Spin excitations in fluctuating stripe phases of doped cuprate superconductors. *Phys. Rev. Lett.* 97(9):097001 (2006).
16. Ikeda Y, *et al.* Transport and thermodynamic studies of stripe and checkerboard ordering in layered nickel oxides $R_{2-x}Sr_xNiO_4$ (R = La and Nd). *J. Phys. Soc. Jpn.* 84(2):023706 (2015).
17. Anissimova S, *et al.* Direct observation of dynamic charge stripes in $La_{2-x}Sr_xNiO_4$. *Nat. Commun.* 5:3467 (2014).
18. Hücker M, *et al.* Unidirectional diagonal order and three-dimensional stacking of charge stripes in orthorhombic $Pr_{1.67}Sr_{0.33}NiO_4$ and $Nd_{1.67}Sr_{0.33}NiO_4$. *Phys. Rev. B* 74(8):085112 (2006).
19. Lee SH, Cheong SW. Melting of quasi-two-dimensional charge stripes in $La_{5/3}Sr_{1/3}NiO_4$. *Phys. Rev. Lett.* 79(13):2514-2517 (1997).
20. Ulbrich H, *et al.* Evidence for charge orbital and spin stripe order in an overdoped manganite. *Phys. Rev. Lett.* 106(15):157201 (2011).
21. Popović Z, Satpathy S. Charge stacking in the half-doped manganites. *J. Appl. Phys.* 91(10):8132-8134 (2002).

22. Sun Z, *et al.* Localization of electrons due to orbitally ordered bi-stripes in the bilayer manganite $La_{2-2x}Sr_{1+2x}Mn_2O_7$ ($x \sim 0.59$). *Proc. Natl. Acad. Sci. USA* 108(29):11799-11803 (2011).
23. Abeykoon AMM, *et al.* Evidence for short-range-ordered charge stripes far above the charge-ordering transition in $La_{1.67}Sr_{0.33}NiO_4$. *Phys. Rev. Lett.* 111(9):096404 (2013).
24. Woo H, *et al.* Mapping spin-wave dispersions in stripe-ordered $La_{2-x}Sr_xNiO_4$ ($x$=0.275, 0.333). *Phys. Rev. B* 72(6):064437 (2005).
25. Yoshizawa H, *et al.* Stripe order at low temperatures in $La_{2-x}Sr_xNiO_4$ with 0.289<=x<=0.5. *Phys. Rev. B* 61(2):R854-R857 (2000).
26. Zachar O, Kivelson SA, Emery VJ. Landau theory of stripe phases in cuprates and nickelates. *Phys. Rev. B* 57(3):1422-1426 (1998).
27. Cheong SW, *et al.* Charge-ordered states in $(La,Sr)_2NiO_4$ for hole concentrations $n_h$=1/3 and 1/2. *Phys. Rev. B* 49(10):7088-7091 (1994).
28. Chen CH, Cheong SW, Cooper AS. Charge modulations in $La_{2-x}Sr_xNiO_{4+y}$ ordering of polarons. *Phys. Rev. Lett.* 71(15):2461-2464 (1993).
29. Anisimov VI, Bukhvalov D, Rice TM. Electronic structure of possible nickelate analogs to the cuprates. *Phys. Rev. B* 59(12):7901-7906 (1999).
30. Poltavets VV, *et al.* $La_3Ni_2O_6$: a new double T'-type nickelate with infinite $Ni^{1+/2+}O_2$ layers. *J. Am. Chem. Soc.* 128(28):9050-9051 (2006).
31. ApRoberts-Warren N, *et al.* NMR evidence for spin fluctuations in the bilayer nickelate $La_3Ni_2O_6$. *Phys. Rev. B* 88(7) (2013).
32. Lacorre P. Passage from T-type to T'-type arrangement by reducing $R_4Ni_3O_{10}$ to $R_4Ni_3O_8$ (R = La, Pr, Nd). *J. Solid State Chem.* 97(2):495-500 (1992).
33. Retoux R, Rodriguez-Carvajal J, Lacorre P. Neutron diffraction and TEM studies of the crystal structure and defects of $Nd_4Ni_3O_8$. *J. Solid State Chem.* 140(2):307-315 (1998).
34. Poltavets VV, *et al.* Crystal structures of $Ln_4Ni_3O_8$ (Ln = La, Nd) triple layer T'-type nickelates. *Inorg. Chem.* 46(25):10887-10891 (2007).
35. Poltavets VV, *et al.* Bulk magnetic order in a two-dimensional $Ni^{1+}/Ni^{2+}$ ($d^9/d^8$) nickelate, isoelectronic with superconducting cuprates. *Phys. Rev. Lett.* 104(20):206403 (2010).
36. ApRoberts-Warren N, *et al.* Critical spin dynamics in the antiferromagnet $La_4Ni_3O_8$ from $^{139}$La nuclear magnetic resonance. *Phys. Rev. B* 83(1):014402 (2011).
37. Cheng JG, *et al.* Pressure effect on the structural transition and suppression of the high-spin state in the triple-layer T'-$La_4Ni_3O_8$. *Phys. Rev. Lett.* 108(23):236403 (2012).
38. Lokshin K, Egami T. Structure and electronic properties of the $La_4Ni_3O_8$. [http://meetings.aps.org/link/BAPS.2011.MAR.T23.11] (2011).
39. Pardo V, Pickett WE. Pressure-induced metal-insulator and spin-state transition in low-valence layered nickelates. *Phys. Rev. B* 85(4):045111 (2012).
40. Hua W. Charge–spin–orbital states in the tri-layered nickelate $La_4Ni_3O_8$: an *ab initio* study. *New J. Phys.* 15(2):023038 (2013).
41. Lee and Pickett argue that the weaker *d-p* hybridization in nickelates renders any such $Cu^{2+}$-$Ni^{1+}$ analogy moot [Lee KW, Pickett WE. Infinite-layer $LaNiO_2$: $Ni^+$ is not $Cu^{2+}$. *Phys. Rev. B* 70(16):165109 (2004)].

42. Ling CD, Argyriou DN, Wu G, Neumeier JJ. Neutron diffraction study of $La_3Ni_2O_7$: structural relationships among $n$=1, 2, and 3 phases $La_{n+1}Ni_nO_{3n+1}$. *J. Solid State Chem.* 152(2):517-525 (2000).
43. Li J, Zhu Y, Tranquada JM, Yamada K, Buttrey DJ. Transmission-electron-microscopy study of charge-stripe order in $La_{1.725}Sr_{0.275}NiO_4$. *Phys. Rev. B* 67(1):012404 (2003).
44. Hayward MA, Green MA, Rosseinsky MJ, Sloan J. Sodium hydride as a powerful reducing agent for topotactic oxide deintercalation: synthesis and characterization of the nickel(I) oxide $LaNiO_2$. *J. Am. Chem. Soc.* 121(38):8843-8854 (1999).
45. Poltavets *et al.* in ref. (34) pointed out that the $La_4Ni_3O_8$ structure does not allow oxygen vacancies despite its preparation by oxygen deintercalation. The data collected on our $La_4Ni_3O_8$ single crystals are in excellent agreement with those reported by Poltavets *et al.* in ref. (35) and Cheng *et al.* in ref. (37), demonstrating that our samples are oxygen stoichiometric.
46. Sugiyama J, Xia C, Tani T. Anisotropic magnetic properties of $Ca_3Co_4O_9$: Evidence for a spin-density-wave transition at 27 K. *Phys. Rev. B* 67(10):104410 (2003).
47. Wooldridge J, Paul DMcK, Balakrishnan G, Lees MR. Investigation of the spin density wave in $Na_xCoO_2$. *J. Phys.: Condens. Matter* 17(4):707 (2005).
48. Klingeler R, Büchner B, Cheong SW, Hücker M. Weak ferromagnetic spin and charge stripe order in $La_{5/3}Sr_{1/3}NiO_4$. *Phys. Rev. B* 72(10):104424 (2005).
49. The curvature feature at ~180 K reported by Cheng *et al.* in ref. (37) is not found in our measurements.
50. We attempted to subtract the $La_3Ni_2O_6$ heat capacity data kindly provided by Dr. K. Lokshin scaled by an appropriate ratio, but the agreement above and below the transition was poor. We have thus phenomenologically modeled the background as described in the text.
51. Pardo V, Pickett WE. Quantum confinement induced molecular correlated insulating state in $La_4Ni_3O_8$ *Phys. Rev. Lett.* 105(26):266402 (2010).
52. Sarkar S, Dasgupta I, Greenblatt M, Saha-Dasgupta T. Electronic and magnetic structures of bilayer $La_3Ni_2O_6$ and trilayer $La_4Ni_3O_8$ nickelates from first principles. *Phys. Rev. B* 84(18):180411 (2011).
53. Liu T, *et al.* Electronic structure and magnetism of $La_4Ni_3O_8$ from first principles. *J. Phys. Condens. Matter* 24(40):405502 (2012).
54. Liu T, *et al.* Dimensionality-induced insulator-metal crossover in layered nickelates $La_{n+1}Ni_nO_{2n+2}$ ($n$ = 2, 3, and $\infty$). *AIP Adv.* 4(4):047132 (2014).

# Supporting Information

**Synchrotron x-ray single crystal diffraction of La$_4$Ni$_3$O$_8$**

Single crystal X-ray diffraction data were collected at room temperature with an APEX2 area detector using synchrotron radiation ($\lambda$=0.41328 Å) at Beamline 15-ID-B at the Advanced Photon Source, Argonne National Laboratory. A single crystal of La$_4$Ni$_3$O$_8$ with dimensions of approximately 5 μm on an edge was attached to the tip of a glass fiber and mounted on the goniometer. Indexing was performed using Bruker APEX2 software(S1). Data integration and cell refinement were performed using SAINT, and multi-scan absorption corrections were applied using the SADABS program. The structure was solved by direct methods and refined with full matrix least-squares methods on $F^2$. All atoms were modeled using anisotropic ADPs, and the refinements converged for $I > 2\sigma(I)$. Calculations were performed using the SHELXTL crystallographic software package(S2). Details of crystal parameters, data collection and structure refinement are summarized in **Table I**. Atomic positions are presented in **Table II**. Further details of the crystal structures may be obtained from Fachinformationszentrum Karlsruhe, 76344 Eggenstein-Leopoldshafen, Germany (fax: (+49) 7247-808-666; E-mail: crysdata@fiz-karlsruhe.de, http://www.fiz-karlsruhe.de/request_for_deposited_data.html) on quoting the appropriate CSD number 429798.

**Table I**. Crystal data and structure refinement for La$_4$Ni$_3$O$_8$ at 296(2) K.

| | | |
|---|---|---|
| Empirical formula | La$_4$Ni$_3$O$_8$ | |
| Formula weight | 859.77 | |
| Temperature | 296(2) K | |
| Wavelength | 0.41328 Å | |
| Crystal system, space group | Tetragonal I4/mmm | Tetragonal F4/mmm |
| Unit cell dimensions | $a$ = 3.9700(5) Å, $\alpha$ = 90° | $a$ = 5.6144 Å, $\alpha$ = 90° |
| | $b$ = 3.9700(5) Å, $\beta$ = 90° | $b$ = 5.6144 Å, $\beta$ = 90° |
| | $c$ = 26.092(3) Å, $\gamma$ = 90° | $c$ = 26.092(3) Å, $\gamma$ = 90° |
| Volume, Z | 411.24(11) Å$^3$, 2 | 822.48 Å$^3$, 4 |
| Density (calculated) | 6.943 g/cm$^3$ | |
| Absorption coefficient | 5.668 mm$^{-1}$ | |
| F(000) | 752 | 1504 |
| $\theta$ range for data collection | 0.908 to 14.402° | |
| Reflections collected/Independent | 1609/128 [$R_{int}$ = 0.0231] | |
| Completeness to $\theta$ = 14.357° | 83.6% | |
| Refinement method | Full-matrix least-squares on F$^2$ | |
| Data / restraints / parameters | 128 / 0 / 21 | |
| Goodness-of-fit | 1.441 | |
| Final R indices [$I > 2\sigma(I)$] | $R_{obs}$ = 0.0170, $wR_{obs}$ = 0.0402 | |
| R indices [all data] | $R_{all}$ = 0.0207, $wR_{all}$ = 0.0517 | |
| Largest diff. peak and hole | 0.921 and -1.046 e·Å$^{-3}$ | |

$R = \Sigma||F_o|-|F_c||/\Sigma|F_o|$, $wR = \{\Sigma[w(|F_o|^2 - |F_c|^2)^2]/\Sigma[w(|F_o|^4)]\}^{1/2}$ and w=1/[$\sigma^2$(Fo$^2$)+10.5182P] where P=(Fo$^2$+2Fc$^2$)/3

**Table II**. Atomic coordinates and equivalent isotropic displacement parameters (Å$^2$) for La$_4$Ni$_3$O$_8$ at 296(2) K with estimated standard deviations in parentheses.

| Label | Wyckoff position | I4/mmm $a=b=3.9700(5)$ Å, $c=26.092(3)$ Å | | | F4/mmm $a=b=5.6144$ Å, $c=26.092$ Å | | | Occupancy | $U_{eq}$* |
|---|---|---|---|---|---|---|---|---|---|
| | | x | y | z | x | y | z | | |
| La(1) | 4e | 0 | 0 | 0.4339(1) | 0 | 0 | 0.4339(1) | 1 | 0.004(1) |
| La(2) | 4e | 0 | 0 | 0.2990(1) | 0 | 0 | 0.2990(1) | 1 | 0.004(1) |
| Ni(1) | 2a | 0 | 0 | 0 | 0 | 0 | 0 | 1 | 0.003(1) |
| Ni(2) | 4e | 0 | 0 | 0.1250(1) | 0 | 0 | 0.1250(1) | 1 | 0.005(1) |
| O(1) | 4c | 0 | 0.5 | 0 | 0.25 | 0.25 | 0 | 1 | 0.007(2) |
| O(2) | 8g | 0 | 0.5 | 0.1261(2) | 0.25 | 0.25 | 0.1261(2) | 1 | 0.007(2) |
| O(3) | 4d | 0 | 0.5 | 0.2500 | 0.25 | 0.25 | 0.2500 | 1 | 0.005(2) |

*$U_{eq}$ is defined as one third of the trace of the orthogonalized $U_{ij}$ tensor.

## Diffuse scattering along *c**

We noticed the presence of forbidden peaks in the *hk*0 plane; here we show that they arise from diffuse scattering attributed to stacking faults along the *c* axis. Fig. S1*A* shows a reconstructed *hk*0 reciprocal space plane of a La$_4$Ni$_3$O$_8$ crystal measured at 95 K using APEX2. The yellow circles emphasize forbidden peaks, such as (5$\bar{5}$0) (#3), (5$\bar{3}$0) (#2), (5$\bar{1}$0) (#1). Fig. S1*B* shows a three-dimensional plot of the rectangular area shown in Fig. S1*A*; one can see clearly the three forbidden peaks (#1, #2, and #3) in addition to three superlattice peaks. Fig. S1*C* shows a reconstructed 5*kl* plane to check whether there is intensity for these forbidden peaks along the *l* direction. Fig. S1*D* shows the peaks (5$\bar{5}$1) and (5$\bar{5}\bar{1}$), and an expanded picture of #3 on the right of Fig. S1*D*. As can be seen, (5$\bar{5}$1) and (5$\bar{5}\bar{1}$) are strongly allowed Bragg peaks, and the tails from these two neighboring peaks extend into the region of (5$\bar{5}$0). We adjusted the sample-detector distance, but could not separate them adequately. Similar behavior is found for each forbidden peak, as shown in Fig. S1*A*. This diffuse rod of intensity is observed as a spot in the *hk*0 plane, but does not represent a true Bragg reflection that would indicate a lower average symmetry. More importantly, the intensity at the forbidden positions is also found above the 105 K transition, and exhibits no temperature dependence. This indicates that the diffuse scattering comes from static disorder in the system unrelated to the charge stripes.

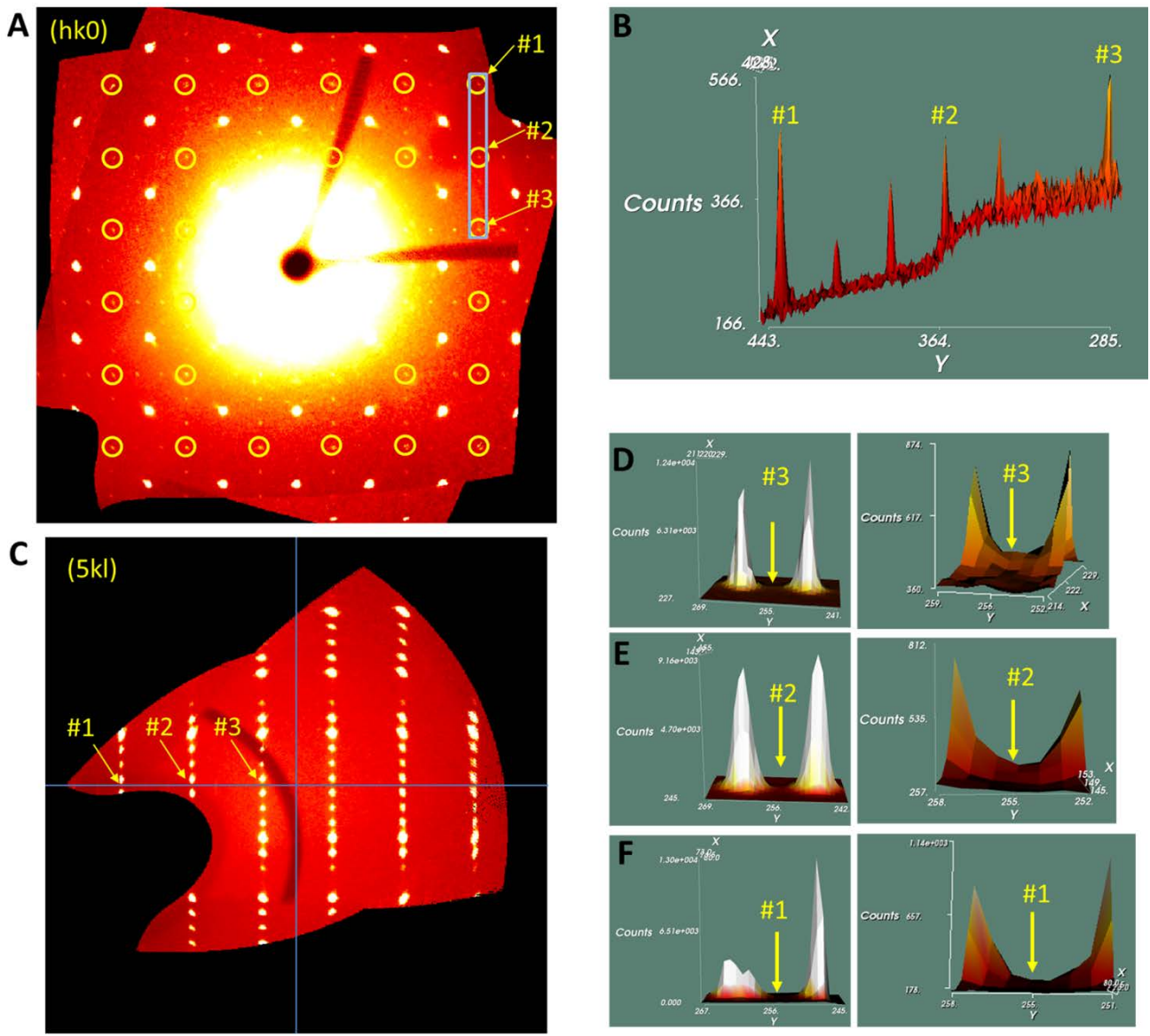

**Fig. S1**. Reconstructed *hk*0 and 5*kl* reciprocal space planes and three dimensional plots of the forbidden peaks (*F*4/*mmm* setting) of a $La_4Ni_3O_8$ crystal measured at 95 K. Note the units of X and Y axes are detector pixels.

**Variable-temperature high-resolution x-ray powder diffraction**

High-resolution x-ray powder diffraction data were collected on pulverized La-438 crystals at beamline 11-BM at APS. The sample was loaded into a $\Phi=0.8$ mm Kapton capillary tube that was spun continuously at 5600 rpm during data collection. Diffraction patterns were recorded on cooling ($\lambda=0.414125$ Å) at 300 K, 250 K, 200-150 K (5 K interval) and 120-82 K (2 K interval). An Oxford Cryostream 700 Plus $N_2$ gas blower was used to control the temperature, and the cooling rate was set to 5 K/min for 250 K and 200-150 K, and 1 K/min for 120-82 K. The temperature was stabilized for 2 min at each set point prior to data collection. Data were analyzed with the Rietveld method using the GSAS(S3) software under the graphical interface EXPGUI(S4). The background at each temperature was fit using a 20-term Chebyshev polynomial (Function #1) and then fixed. Refined parameters include scale factor, lattice parameters, atomic positions, isotropic atomic displacement parameters ($U_{iso}$, all atoms are grouped together), and profile shape parameters. Pseudo-Voigt functions with anisotropic microstrain broadening (function #4)(S5) were used for the peak profiles. For the three-phase refinement of room temperature data, the corresponding profile parameters for each phase were constrained to be the same. The refinement of room-temperature data converged at $R_{wp}=12.51\%$, $R_p=9.25\%$ and $\chi^2=6.74$ with weight fractions of 98.7%, 0.6%, and 0.7% for La-438, $La_3Ni_2O_{6.35}$ and $La_2O_3$, respectively. The low temperature data were refined using a single-phase model.

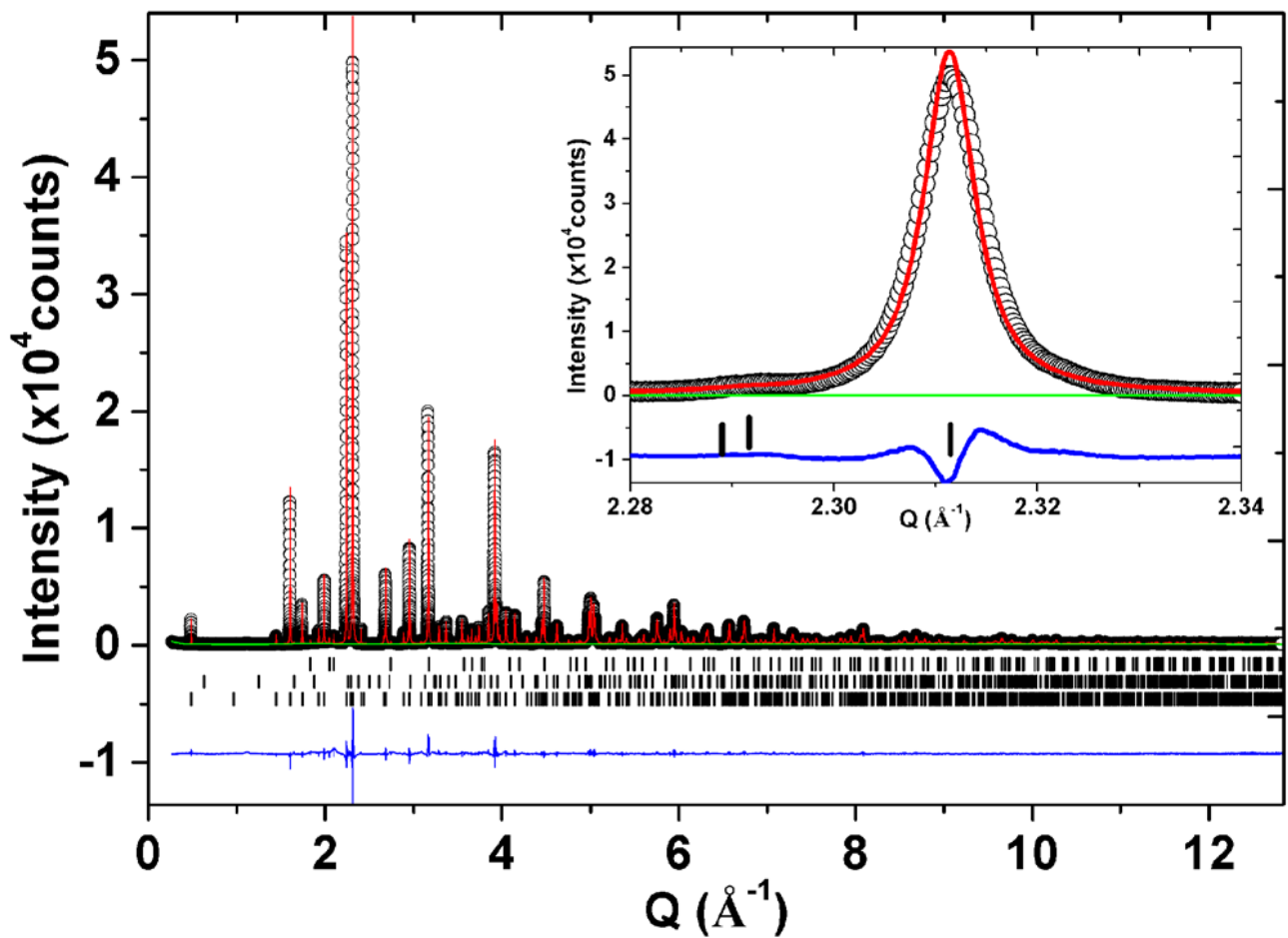

**Fig. S2**. High-resolution synchrotron X-ray pattern at room temperature. The black circles, red curve, green curve, black bars and blue curve correspond to the observed data, calculated intensity, background, Bragg peaks, and difference curve, respectively. Note for the black bars, $La_4Ni_3O_8$ (bottom), $La_3Ni_2O_{6.35}$(middle), and $La_2O_3$(top).

# Charge stripe stacking models

To understand the charge stripe arrangement in $La_4Ni_3O_8$, we studied several models. While there could be other possibilities, we list here those we consider most reasonable.

**Model #1**: This model has in-phase stacking within each trilayer as presented in the main text (see main text, Fig. 3*J*). It is characterized by two sets of superlattice reflections (weaker and stronger) around each of the main Bragg peaks in the $hk0$ plane, and indeed both sets are observed in the data shown in Fig. 3*A*. Importantly, maxima in the structure factors are predicted for $l=8n$, where $n$ is an integer, in accordance with the observations shown in Fig. 3*C*. Among those tested, this model appears to be the best description for $La_4Ni_3O_8$.

**Model #2**: In this model, shown in Fig. S3, the charges are not stacked in-phase within each trilayer. Instead, the charge in the center layer of each trilayer is shifted along *a* with respect to the top and bottom layers within the same trilayer. This generates sets of superlattice maxima centered at $l=4+8n$, where $n$ is an integer, which is at odds with the observations of maxima at $l=8n$.

**Model #3**: In this model, shown in Fig. S4, the top, center, and bottom layers within each trilayer possess stripes that are shifted with respect to one another. This model differs from observation because it generates zero intensity for superlattice reflections in the $hk0$ plane.

**Model #4**: In this model, shown in Fig. S5, charges are stacked in-phase within each trilayer like in Model #1. The trilayer containing the $z=0$ plane is exactly the same as its counterpart in Model #1; however, the charges in the $z=0.5$ layer are shifted along *a* by one Ni-Ni distance with respect to that in Model #1. In this way, the stripes in the $z=0$ and $z=0.5$ trilayers appear stacked when a projection is viewed along *b*. This model does not reproduce all of the observed superlattice reflections in the $hk0$ plane. Specifically, it generates only four superlattice peaks around each main Bragg reflection as shown in Fig. S5.

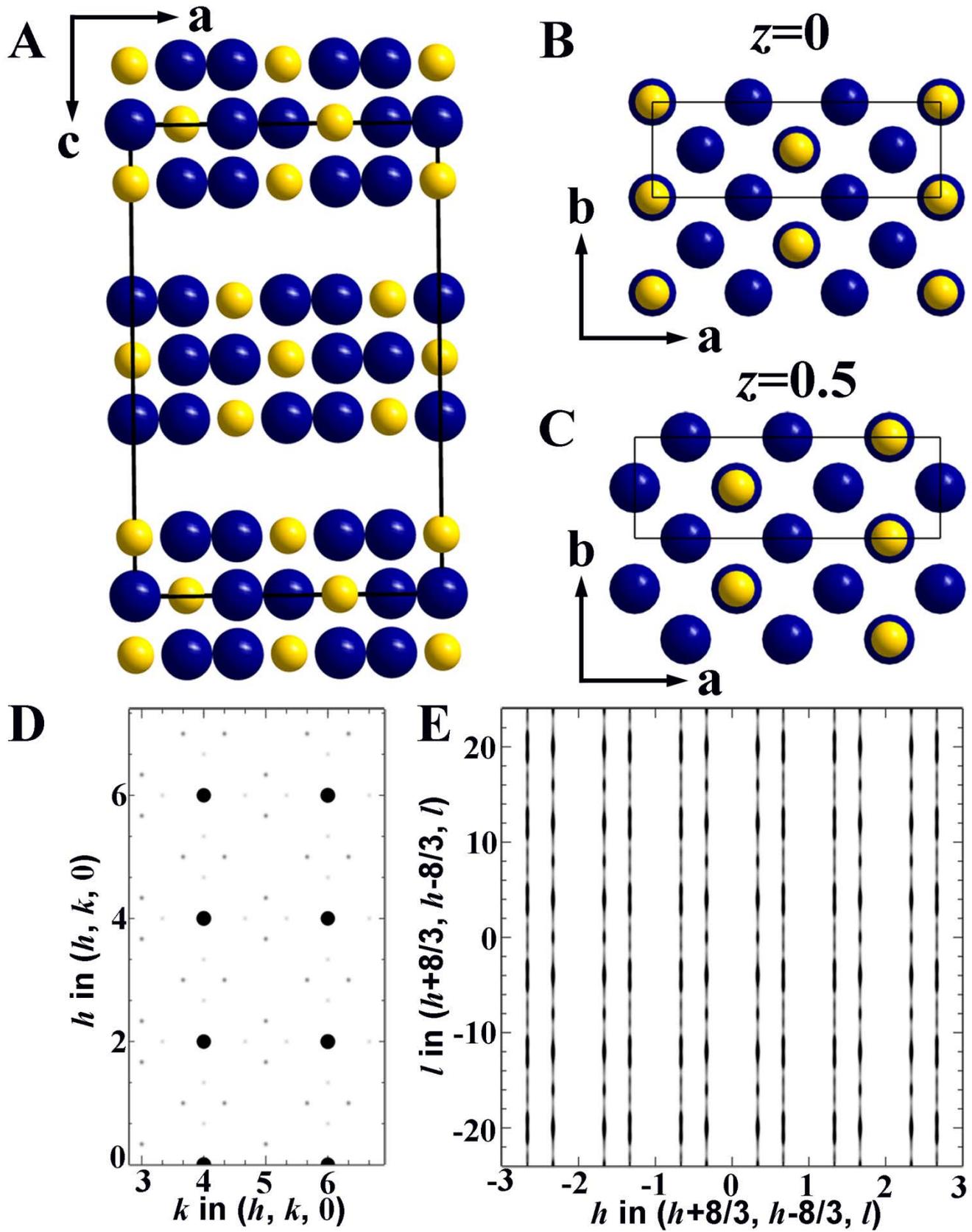

**Fig. S3** Charge stripe model #2 and simulated diffraction patterns.

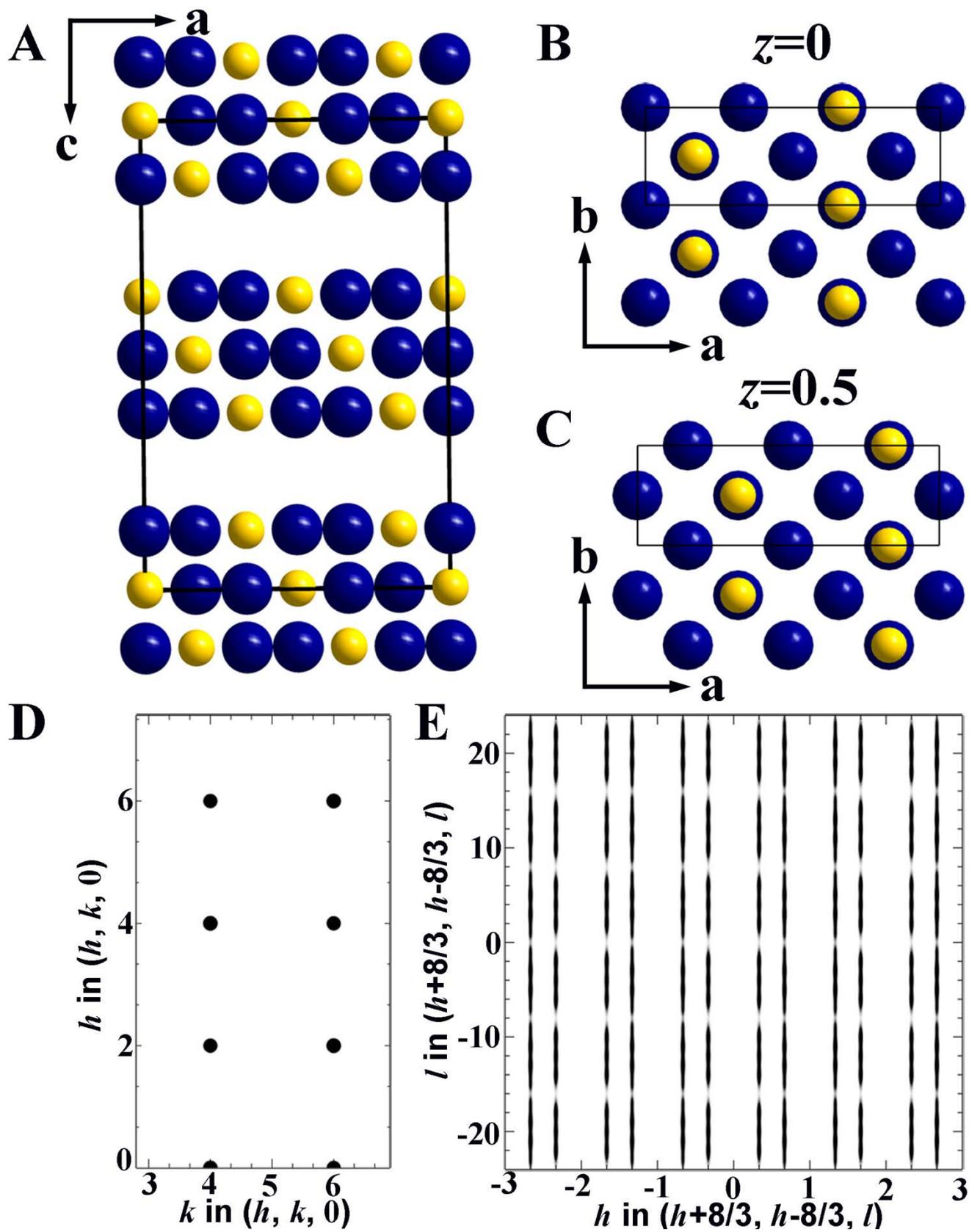

**Fig. S4** Charge stripe model #3 and simulated diffraction patterns.

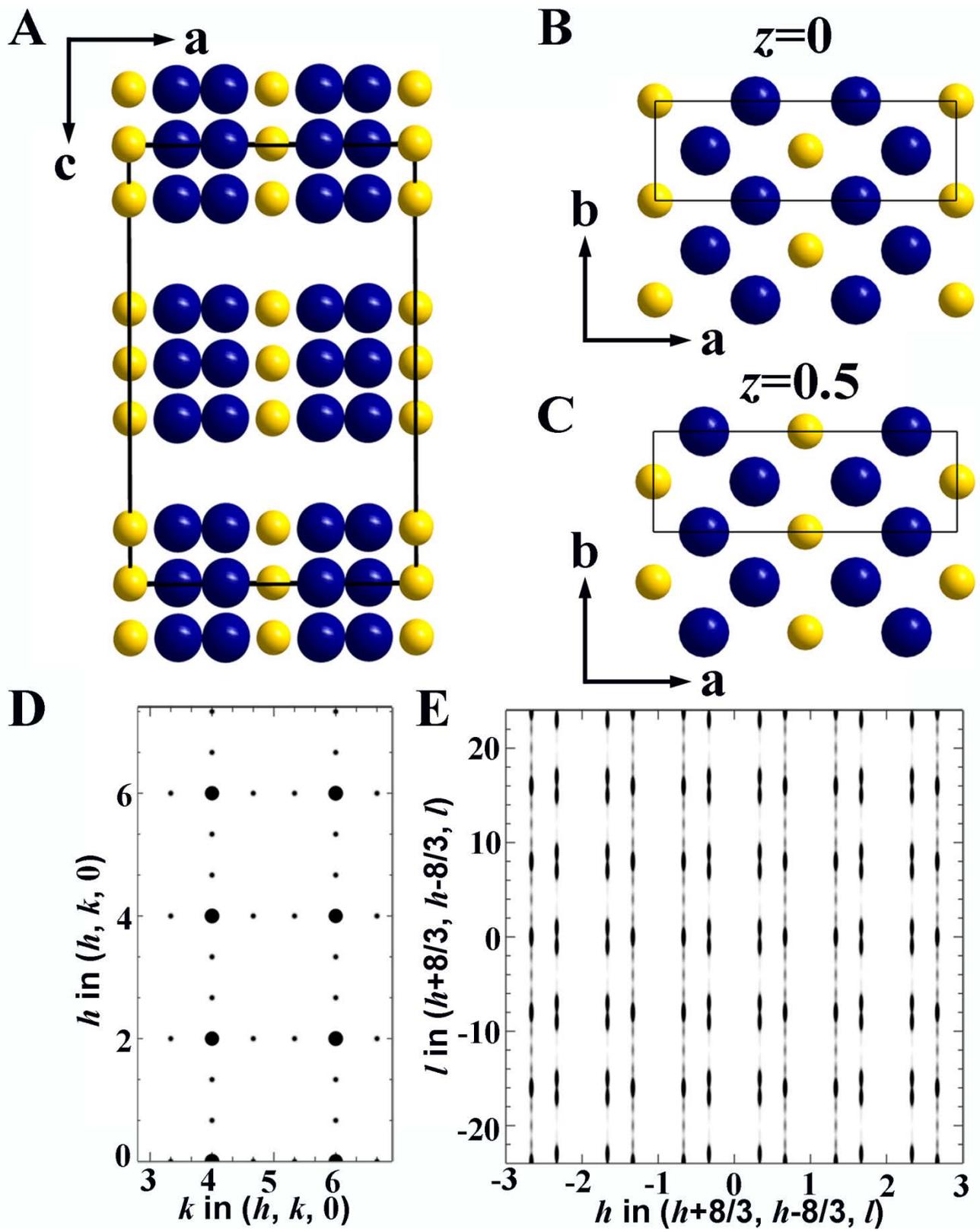

**Fig. S5** Charge stripe model #4 and simulated diffraction patterns.

# References

S1. Bruker APEX2. Bruker Analytical X-ray Instruments, Inc. Madison, Wisconsin, USA (2005).

S2. Sheldrick GM. computer code SHELXTL-version 6.12. Bruker Analytical X-ray Instruments, Inc. Madison, WI (2001).

S3. Larson AC, Dreele RBV. *General Structure Analysis System (GSAS).* Los Alamos National Laboratory Report LAUR 86-748 (2004).

S4. Toby BH. EXPGUI, a graphical user interface for GSAS. *J. Appl. Crystallogr.* 34:210-213 (2001).

S5. Stephens PW. Phenomenological model of anisotropic peak broadening in powder diffraction. *J. Appl. Crystallogr.* 32(2):281-289 (1999).